\documentclass[11pt,reprint,showpacs]{revtex4-1}
\usepackage[usenames]{color}
\usepackage{graphics,amsmath}
\usepackage{epsfig}
\usepackage{multirow}
\usepackage{colordvi}
\begin{document}
\title{ Kinetic instabilities that limit $\beta$ in the edge of a tokamak plasma: a picture of an H-mode pedestal}
\author{D Dickinson$^{1,2}$, C M Roach$^{1}$, S Saarelma$^{1}$, R Scannell$^{1}$, A Kirk$^{1}$ and H R Wilson$^{2}$}
\affiliation{$^1$ EURATOM/CCFE Fusion Association, Culham Science Centre, Abingdon, Oxfordshire, OX14 3DB, UK \\
$^{2}$ York Plasma Institute, Dep't of Physics, University of York, Heslington, York, YO10 5DD, UK}
\date{March 16, 2012}
\begin{abstract}
Plasma equilibria reconstructed from the Mega-Amp Spherical Tokamak (MAST) have 
 sufﬁcient resolution to capture plasma evolution during the short period between 
 edge-localized modes (ELMs). Immediately after the ELM steep gradients in pressure, $P$, 
 and density, $n_e$, form pedestals close to the separatrix, and they then
 expand into the core. Local gyrokinetic analysis over the ELM cycle reveals the dominant 
 microinstabilities at perpendicular wavelengths of the order of the 
 ion Larmor radius. These are kinetic ballooning modes (KBMs) in the pedestal 
 and microtearing modes (MTMs) in the core close to the pedestal top. 
 The evolving growth rate spectra, supported by gyrokinetic analysis using artificial 
 local equilibrium scans, suggest a new physical picture for the formation and arrest 
 of this pedestal.
\end{abstract}

\vspace*{-2mm}
\pacs{52.25.Fi, 52.30.Gz, 52.35.Lv, 52.35.Qz, 52.55.Fa, 52.65.Tt}
\vspace*{-3mm}
\maketitle

\noindent
{\em Introduction:--} In the high confinement mode (H-mode) of operation in tokamaks \cite{WAGNER_PRL1982}, containment of magnetised plasma is enhanced through the formation of a narrow insulating layer that supports steep gradients of density $n_e$ and/or temperature, at the edge of the plasma. A steep pressure pedestal then grows until it either saturates, or it is rapidly destroyed by an edge-localised-mode (ELM) \cite{CONNOR_PPCF1998}. Pedestal growth and destruction by the ELM usually occurs as a cyclical process in H-mode plasmas. Pedestal properties are important in determining the capacity for plasma devices to confine energy: e.g. predictions of fusion power in the next step tokamak ITER, based on models of turbulent transport in the core, are sensitive to the size of the edge temperature pedestal \cite{WAKATANI_IPBCH2_NF1999, DOYLE_PIPBCH2_NF2007}.

Analysis of data from many tokamaks, often guided by ideal MHD at infinite toroidal mode number, $n$, is consistent with the hypothesis that kinetic ballooning modes (KBMs) limit the pressure gradient, $dP/dr$, in the pedestal \cite{SNYDER_NF2009,KIRK_PPCF2009,BEURSKENS_POP2011,DICKINSON_PPCF2011}. This is a key ingredient in the pedestal model called EPED \cite{SNYDER_POP2009}, which proposes: drift wave turbulence is suppressed in the pedestal by sheared flow; KBMs constrain the pedestal $dP/dr$ ; and the pedestal width broadens as the edge current density rises, to trigger a finite $n$ ideal MHD peeling-ballooning mode corresponding to the ELM \cite{CONNOR_POP1998}. EPED predictions of pressure pedestal height and width prior to the ELM crash, are consistent with data from several tokamaks \cite{SNYDER_NF2009,SNYDER_POP2009}, and partially supported by recent data \cite{DICKINSON_PPCF2011} from the spherical tokamak MAST \cite{SYKES_POP2001}. The EPED model, however, cannot fully explain the evolution of kinetic profiles.

This Letter presents detailed kinetic analyses of an evolving pedestal equilibrium from MAST \cite{DICKINSON_PPCF2011}, and probes whether microinstability physics helps explain: (i) the inwards expansion of steep pedestal profile gradients into the core, and (ii) mechanisms that limit the fully developed pedestal.

\begin{figure}[ht]
\setlength{\unitlength}{1cm}
\vspace*{-0.5cm}
\includegraphics[scale=0.39]{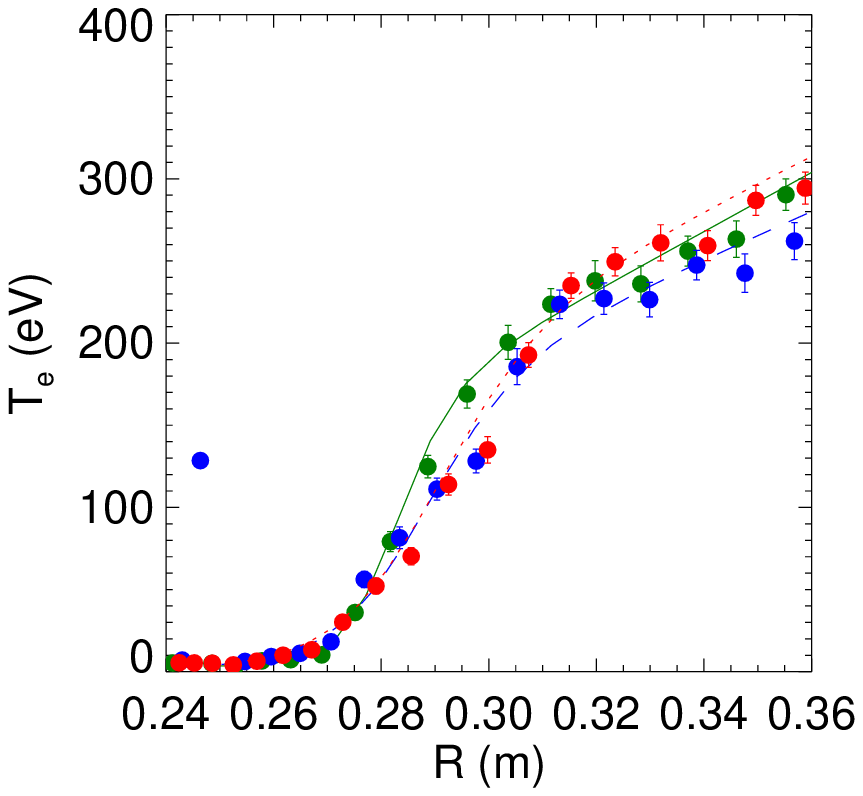}
\put(-3.5,0.){(a)}
\put(0.,0.){(b)}
\includegraphics[scale=0.39]{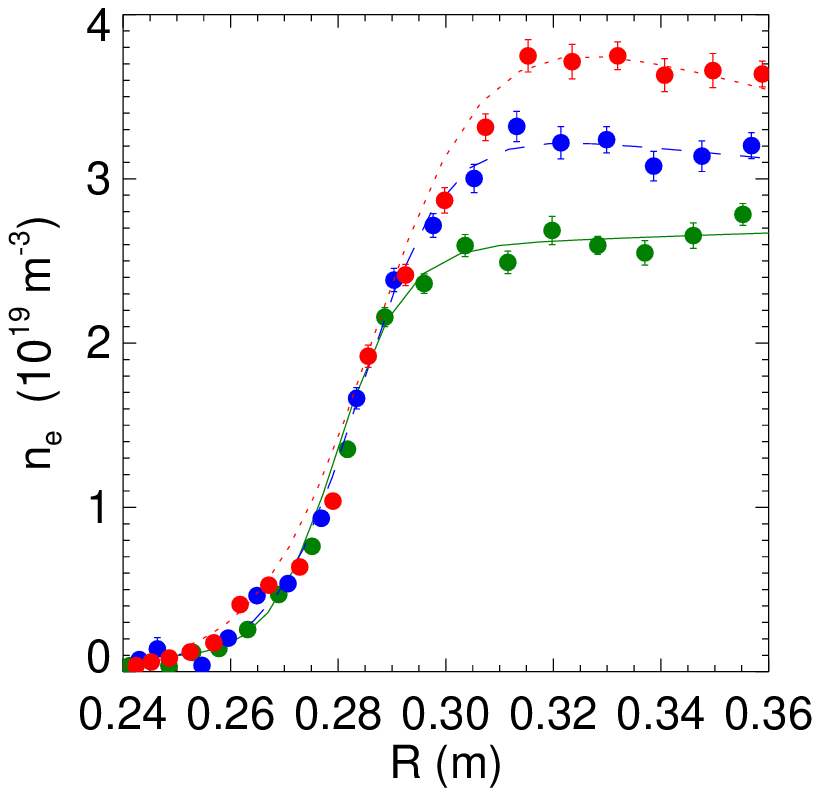} \\
\vspace*{-3mm}
\includegraphics[scale=0.39]{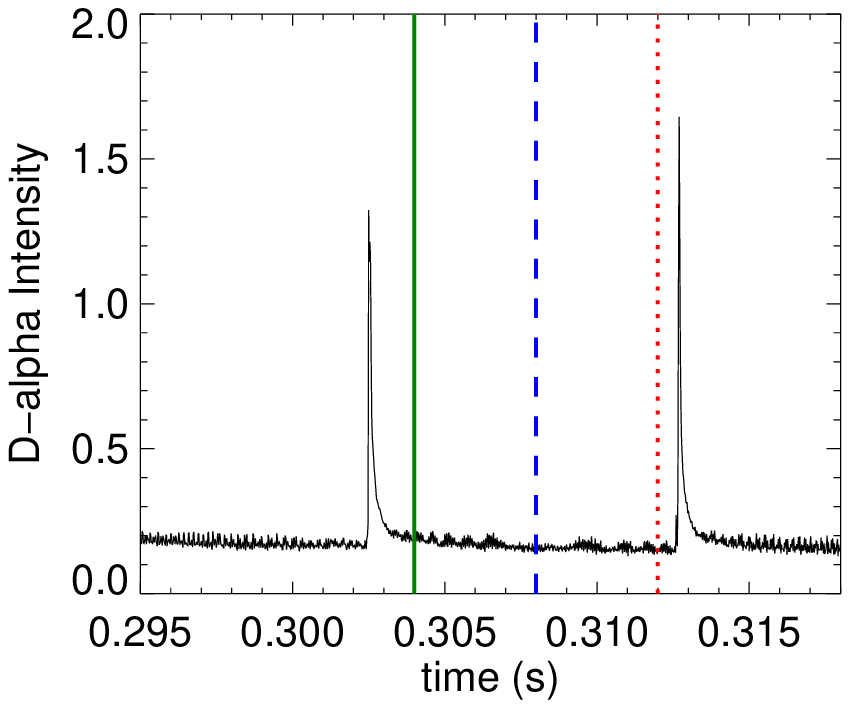}
\put(-4.,0.){(c)}
\put(0.,0.){(d)}
\vspace*{-3mm}
\includegraphics[scale=0.2]{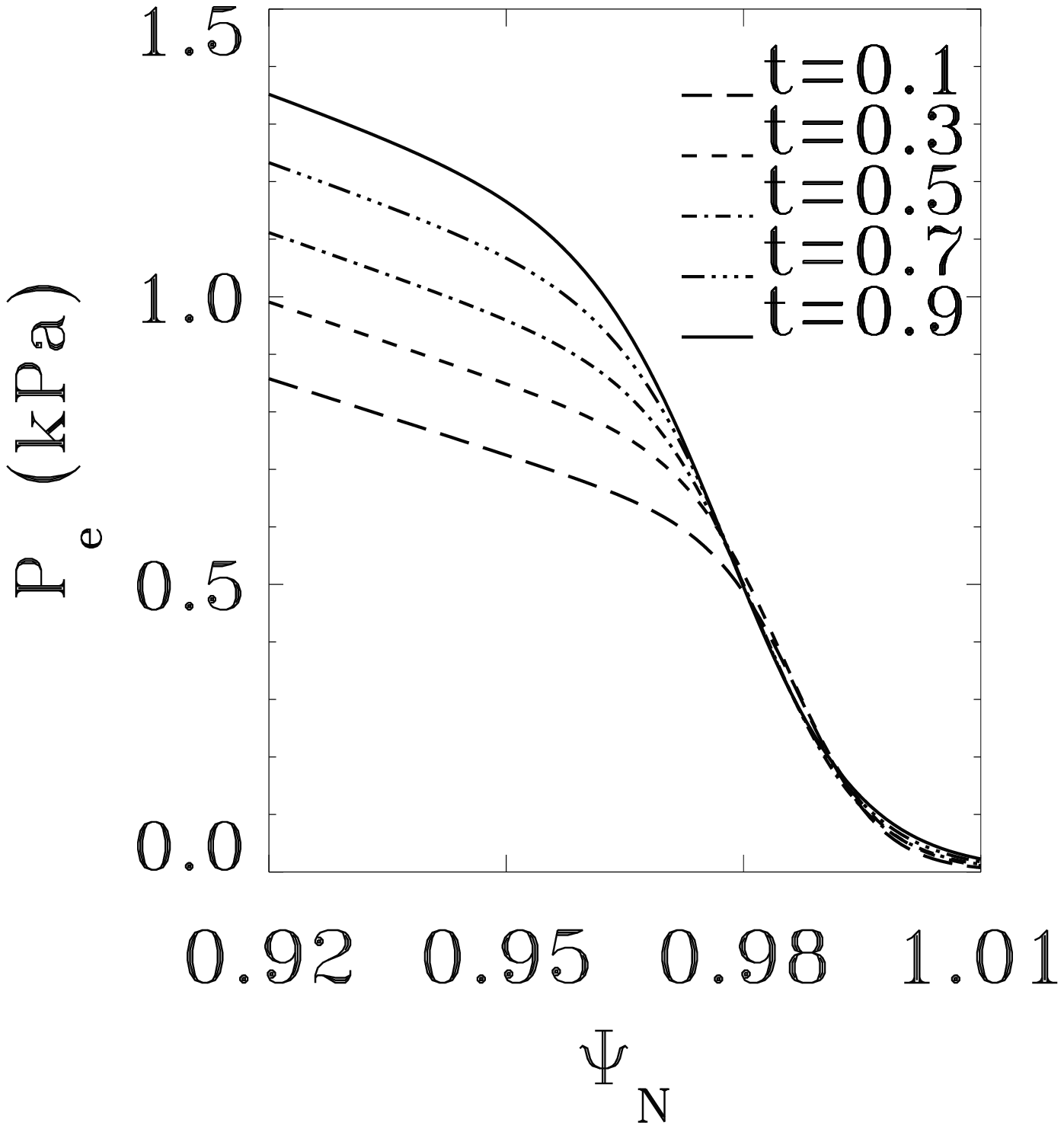}
\caption{\label{fig:TS}TS profiles of (a) $T_e$ and (b) $n_e$ from three times in one inter-ELM period that are indicated on the D-alpha trace (c). TS data from periodic ELMs were used to obtain (d) $P_e(\Psi_N,t)$ over the ELM cycle.}\label{fig:TSdata}
\vspace*{-2mm}
\end{figure}

\noindent
{\em  Inter-ELM Pedestal Profile Evolution in MAST:--}
Data were taken from reproducible MAST H-mode discharges with regular type I ELMs, 
and heated by 3.4MW of neutral beam injection \cite{DICKINSON_PPCF2011}. Profiles of  electron density, $n_e$, and electron temperature, $T_e$, were measured every 4.2ms using a Thomson scattering (TS) system with 130 spatial points \cite{SCANNELL_RSI2010}. Spatial resolution was sufficient to resolve the steep pedestal profiles on the high field side. Figure~\ref{fig:TS}(a-c) shows TS profiles from one ELM cycle that demonstrate stronger changes in $n_e$ than $T_e$, with the density pedestal growing throughout the cycle. 

Time resolution of the ELM cycle was enhanced by ordering TS profiles, from the periodic ELM phases of the discharges, in time with respect to the previous ELM. 
This revealed that $dP_e/dr$ recovers very rapidly close to the separatrix, during the first 10\% of the cycle after the ELM crash.
These profiles were used, assuming ion temperature, $T_i = T_e$, to reconstruct equilibria (see \cite{DICKINSON_PPCF2011} for full details), accommodating the edge bootstrap current \cite{SAUTER_POP2002}, at 
five times normalised to the inter-ELM period, $t$, during one ELM cycle. 
Figure~{\ref{fig:TS}}(d) shows how the edge electron pressure, $P_e(\Psi_N)$, evolves (where $\Psi_N$ is normalised poloidal flux): changes in $n_e$ dominate $P_e$ evolution; the $P_e$ pedestal top moves inwards from $\Psi_N \sim 0.98$ at $t=0.1$ to $\Psi_N \sim 0.96$ at $t=0.9$; and the edge region of high $dP_e/dr$ (i.e. the edge transport barrier) expands into the core with modest change in the peak $dP_e/dr$.  Similar observations during regular type I ELMs have been reported by the tokamak experiments DIII-D \cite{GROEBNER_NF2010} and ASDEX Upgrade \cite{BURCKHART_PPCF2010}.

\noindent
{\em Microstability during Pedestal Evolution:--}
The H-mode pedestal is associated with a strong pressure gradient, which can drive electromagnetic instabilities, 
like the KBM, where the magnetic perturbations are crucial to the instability mechanism. 
Collisionless KBMs and ideal MHD infinite-$n$ ballooning modes are described by related equations and have similar character \cite{TANG_NF1980}.
Finite Larmor radius (FLR) effects, only included in the kinetic treatment, are often stabilising \cite{TANG_NF1981}: the region unstable to 
infinite-$n$ ideal MHD ballooning modes can be broader than the region unstable to KBMs \cite{TANG_NF1981}. 
The stability of infinite-$n$ ballooning modes is often, nevertheless, found to be a reasonable proxy 
for KBM stability \cite{DICKINSON_PPCF2011}. 
The gyrokinetic equation is derived from a first order expansion of the Vlasov equation in the small parameter $\delta= \rho/L$ 
(where $L$ is a typical equilibrium gradient scalelength). Short equilibrium gradient scalelengths increase $\delta$ in the pedestal: 
for ions $\delta_i \sim 0.3$ in the steepest part of the MAST pedestal, but $\delta_i$ is an order of magnitude smaller immediately inside the fully developed pedestal top and 
for electrons $\delta_e \ll 1$ throughout the plasma.  Ideal MHD is applied routinely to analyse pedestal equilibria. 
While gyrokinetics is less accurate in the pedestal than in the core, it improves on MHD in the high $n$ limit through the inclusion of FLR corrections and resonances. Gyrokinetics 
was previously used to study conditions where electron drift waves may cause transport in a simple slab geometry model of the H-mode pedestal \cite{ROGERS_POP2005}.

\vspace*{-0.02mm}
Local gyrokinetic analysis, using the initial value flux-tube code GS2 \cite{KOTSCHENREUTHER_CPC1995}, has obtained the 
microstability properties of the MAST edge plasma during the ELM cycle. GS2 solves the electromagnetic gyrokinetic equation
\cite{ANTONSEN_PF1980} for each plasma species, to find the fastest growing microinstability and its growth rate, $\gamma$, for specified values of 
$k_y \rho_i$ (where $k_y$ is the in-flux-surface component of the perpedicular wavenumber). Dominant growth rate spectra covering $0.05 < k_y \rho_i < 5.5$ 
were computed on 12 surfaces spanning $0.94 < \Psi_N < 0.995$, for 5 times during the ELM cycle, $t=(0.1, 0.3, 0.5, 0.7, 0.9)$. Sheared toroidal flow 
was neglected in this first analysis, and is not expected to be important just inside the pedestal top.

\begin{figure}
\begin{center}
\includegraphics[scale=0.2]{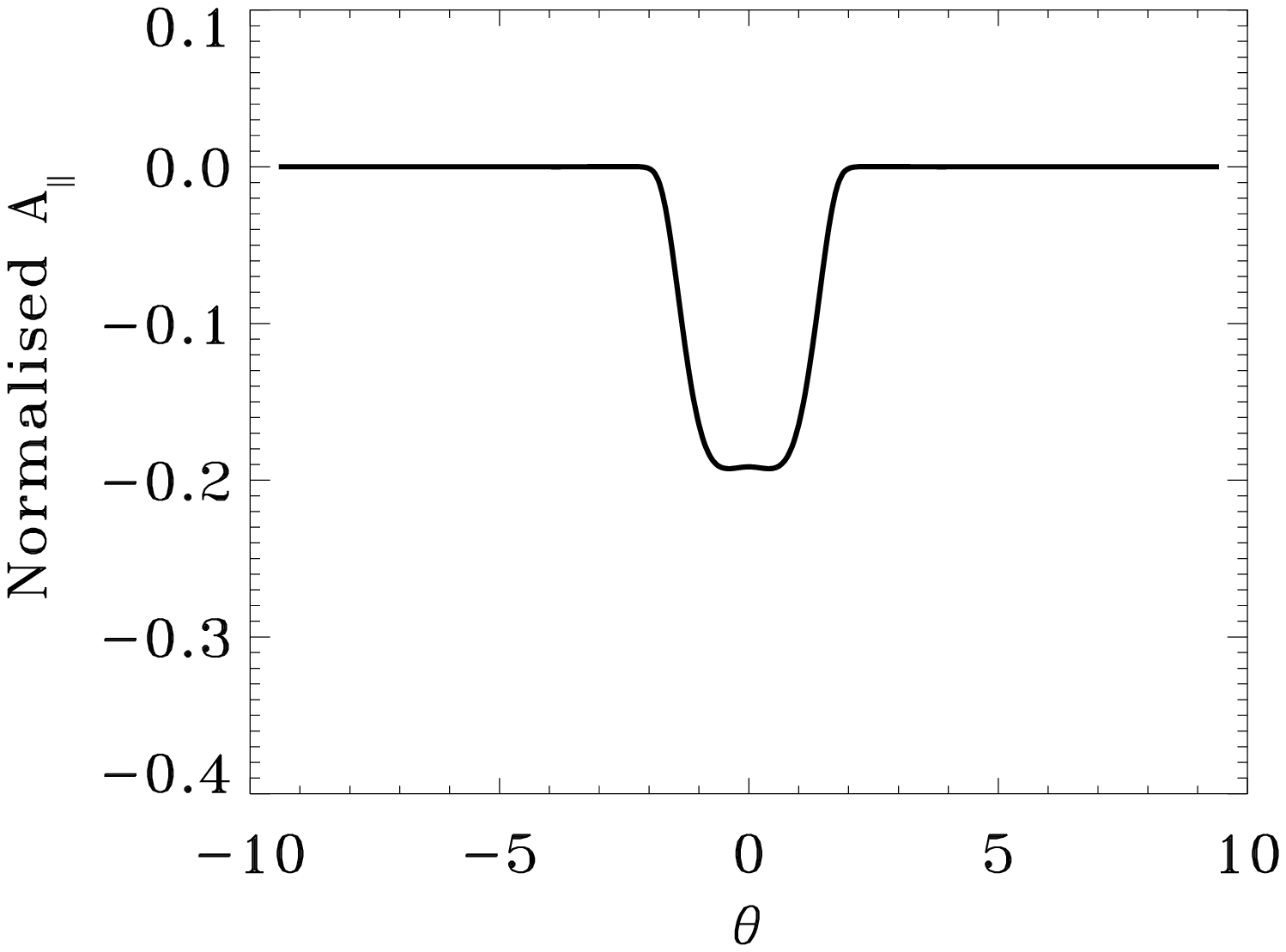}
\includegraphics[scale=0.2]{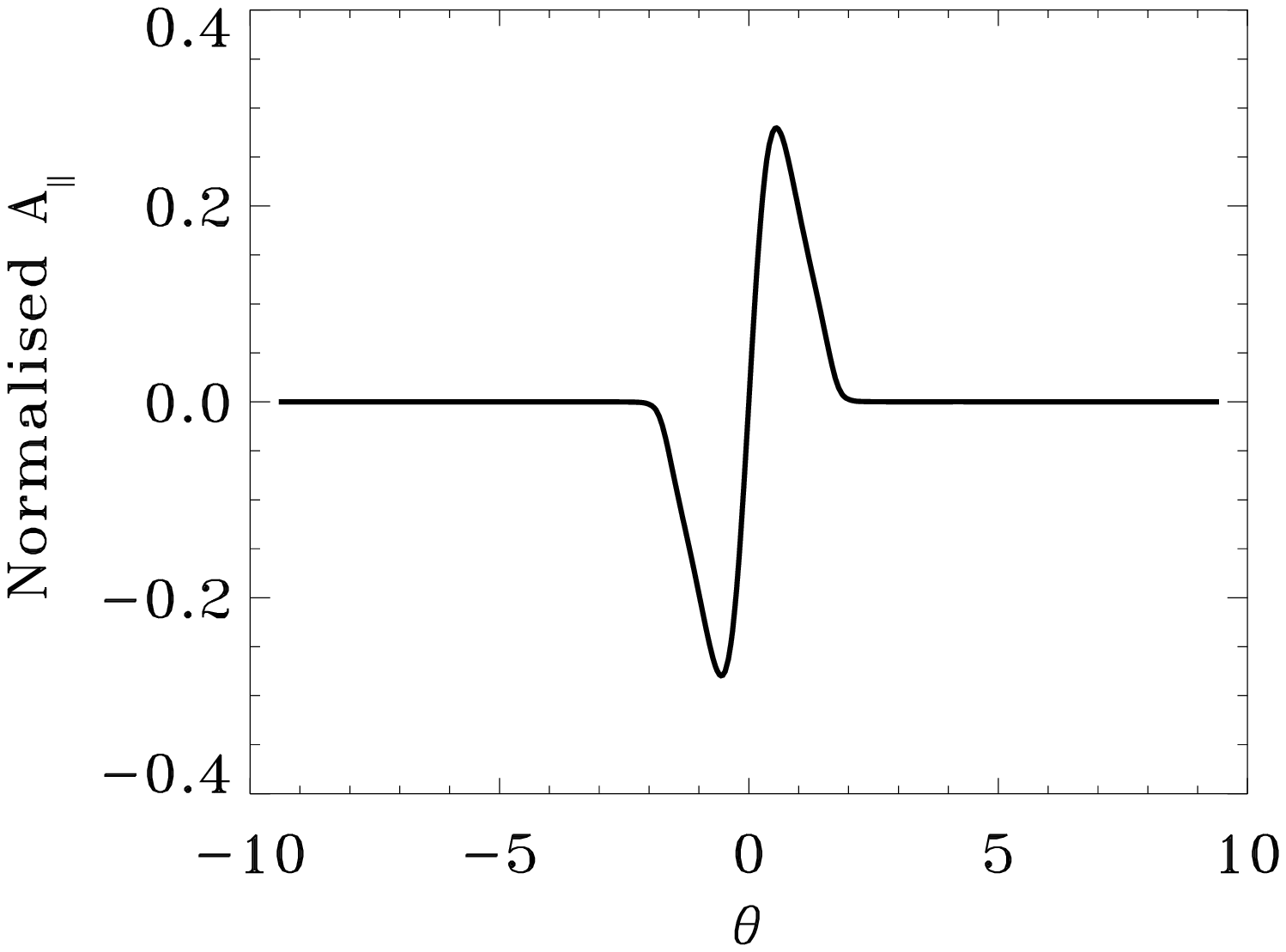}
\end{center}
\vspace*{-0.8cm}
\caption{Real part of the dominant eigenfunctions of $A_{\parallel}$ at $t=0.5$  plotted as functions of ballooning angle, $\theta$, exhibiting (a) tearing parity at $k_y\rho_i=3.28$, $\Psi_N=0.95$ and (b) twisting parity at $k_y\rho_i=0.218$, $\Psi_N=0.98$.}
\label{fig:Tear_KBM_Efunc}
\vspace*{-2mm}
\end{figure}

In the steep $dP/dr$ pedestal region, the dominant microinstabilities grow around $k_y \rho_i \sim 0.2$ and are KBMs with twisting parity. There is a sharp transition to microtearing modes (MTMs) growing around $k_y \rho_i \sim 3$ in the shallower gradient region immediately inside the top of the pedestal. The evolution of the radial region where KBMs dominate is similar to the region found to be unstable to infinite-$n$ ideal MHD ballooning modes \cite{DICKINSON_PPCF2011}. Figure~{\ref{fig:Tear_KBM_Efunc}} illustrates typical eigenfunctions of the parallel magnetic vector potential, $A_{\parallel}$, for MTMs and KBMs.
KBMs are driven by both temperature and density gradients, propagate in the ion diamagnetic drift direction, and typically have a ratio of electron collision frequency to mode frequency $\nu_e/\omega > 1$. MTMs are driven by the electron temperature gradient at large $\eta_e=L_{n_e}/L_{T_e}$, propagate in the electron diamagnetic drift direction, and have electron collisionality 
$\nu_e/\omega \sim O(1)$. The MTMs have similar characteristics to  modes reported at mid-radius in spherical tokamak plasmas 
{\cite{APPLEGATE_PPCF2007,APPLEGATE_POP2004,WONG_POP2008}}, with two notable exceptions: the edge modes here (i) are considerably less extended in ballooning angle, $\theta$, and (ii) arise at 
substantially higher $k_y \rho_i$.  
Microtearing modes have also previously been reported to be unstable  close to the edge in ASDEX Upgrade  {\cite{TOLD_POP2008}}, and in conceptual spherical tokamak burning plasma devices \cite{KOTSCHENREUTHER_NF2000,WILSON_NF2004}. 
Recent measurements from the pedestal region between ELMs in DIII-D {\cite{YAN_POP2011}} find high and low $\omega$ bands of turbulence propagating in the electron and ion drift directions respectively, which are consistent with the properties of the MTMs and KBMs reported here.

\begin{figure}[h] 
\setlength{\unitlength}{1.0cm}
\vspace*{-0.8cm}
\begin{center}
\hspace*{-0.9cm} \includegraphics[scale=0.3]{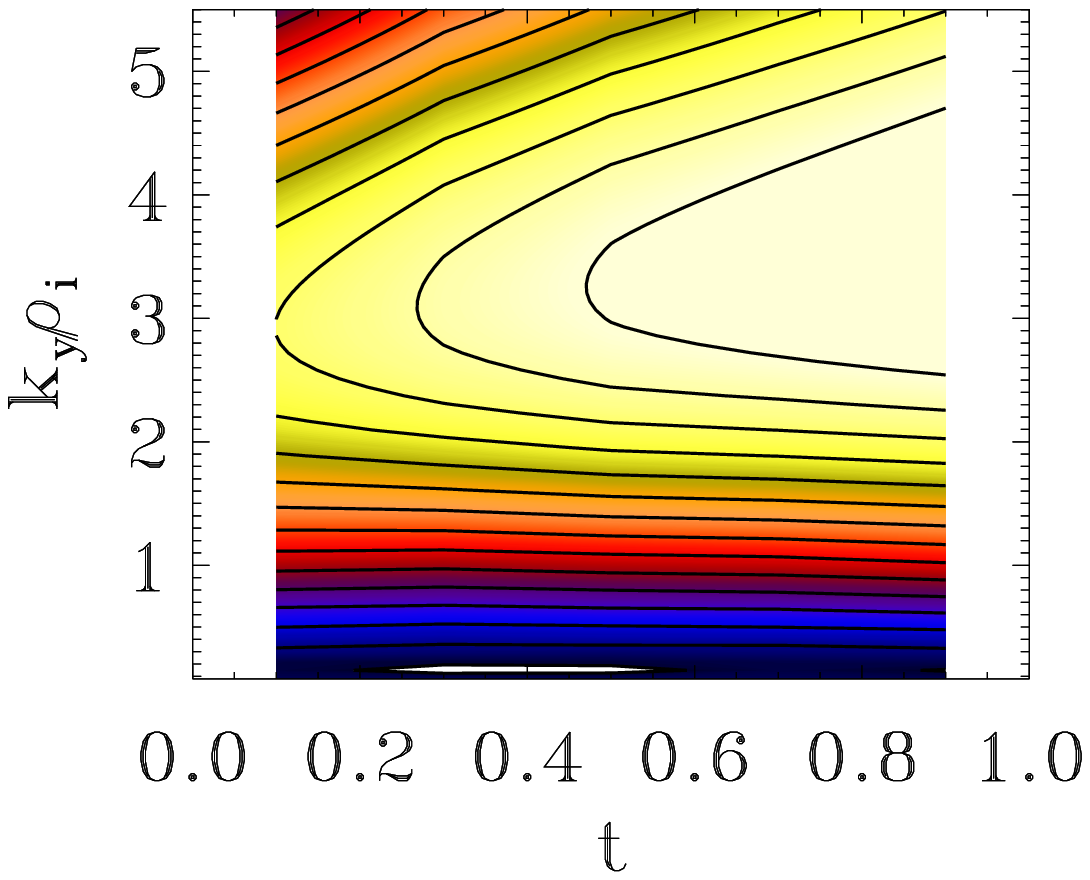} 
\put(-2.7,1.8){\color{black} \bf \tiny MTM}
\hspace*{-1.3cm} \put(-3.4,0.1){(a)} \put(0.5,0.1){(b)}\includegraphics[scale=0.3]{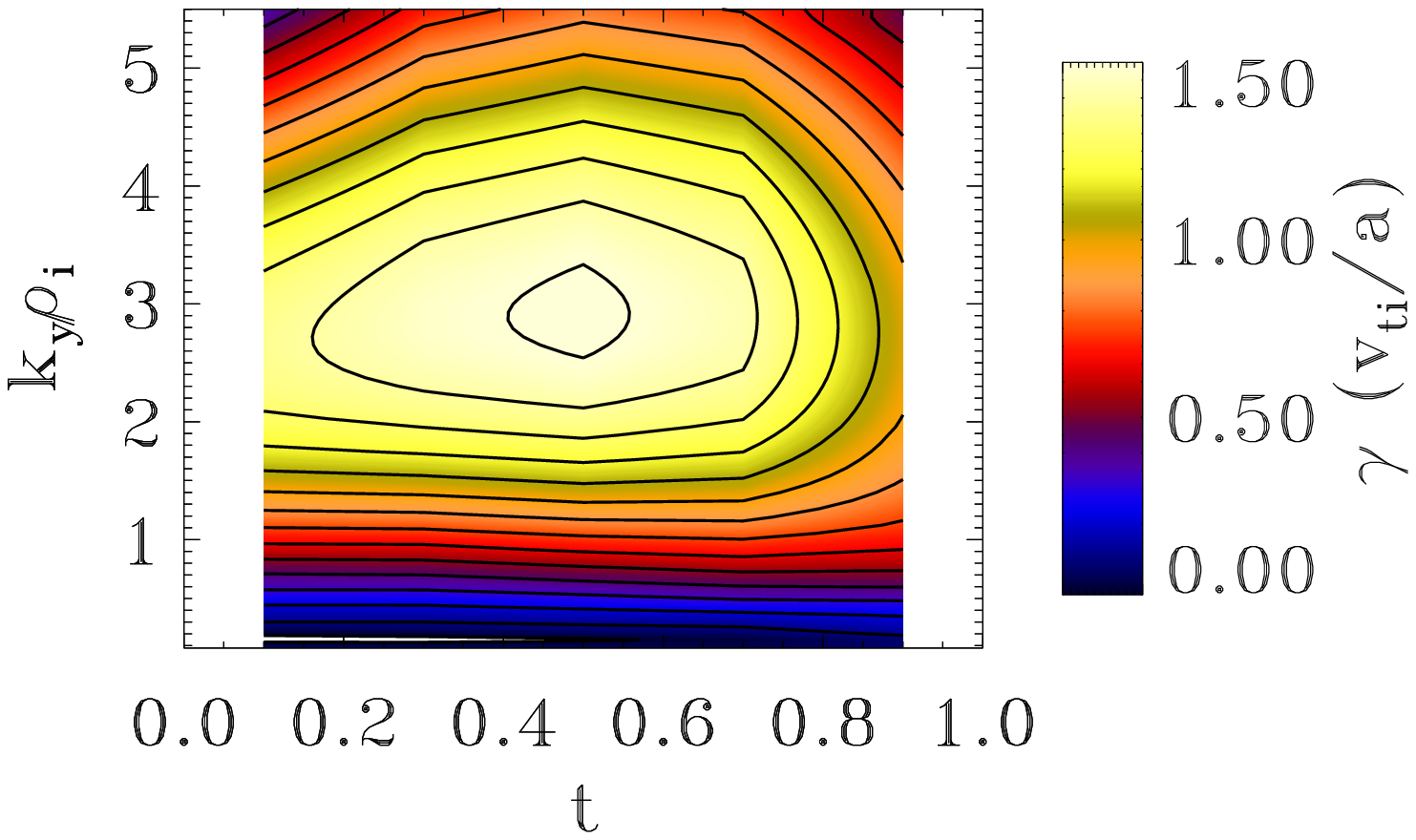} 
\put(-3.0,1.7){\color{black} \bf \tiny MTM}\\
\vspace*{-0.6cm}
\hspace*{-0.9cm} \includegraphics[scale=0.3]{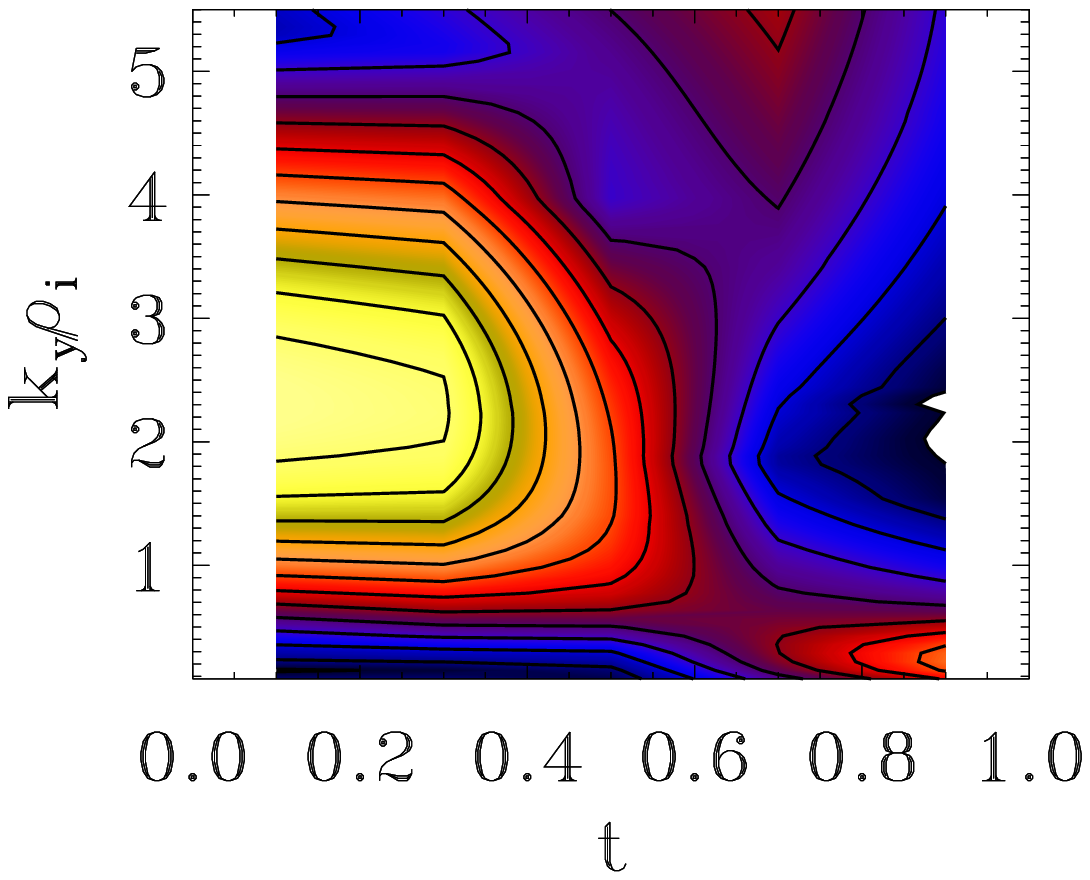} 
\put(-3.8,1.5){\color{black} \bf \tiny MTM}
\put(-2.5,0.8){\color{white} \bf \tiny KBM}
\hspace*{-1.3cm} \put(-3.4,0.1){(c)} \put(0.5,0.1){(d)} \includegraphics[scale=0.3]{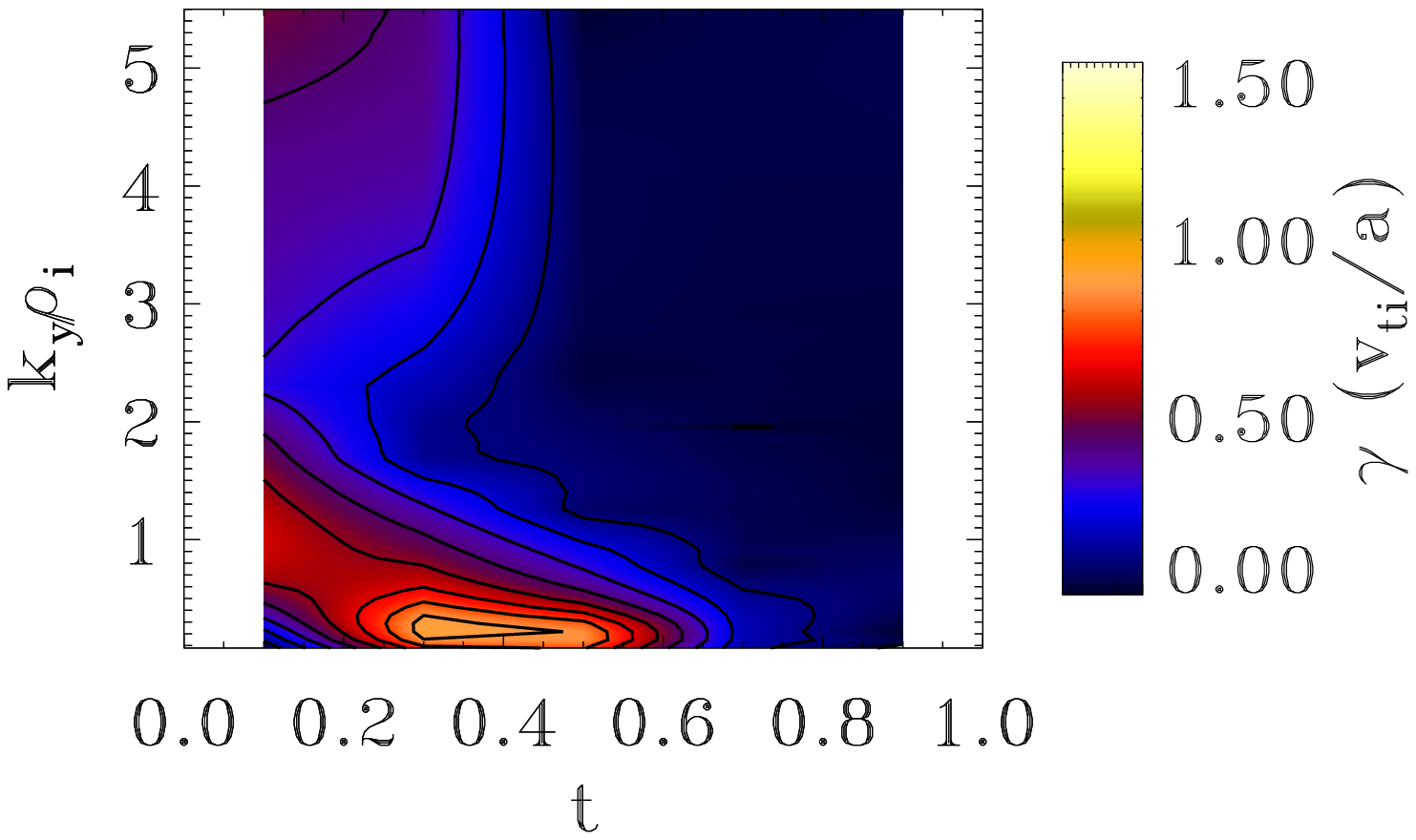} 
\put(-3.,0.9){\color{white} \bf \tiny KBM}
\end{center}
\vspace*{-0.7cm}
\caption{
$\gamma(k_y \rho_i, t)$ over the ELM cycle at $\Psi_N=0.95, 0.96, 0.97, 0.98$ in (a)-(d) respectively. $\gamma$ is normalised to $v_{ti}/a$, $v_{ti}$ is the local ion thermal velocity and $a$ is the minor radius.}
\label{fig:specevol}
\vspace*{-2mm}
\end{figure}

Figure~\ref{fig:specevol} shows the ELM cycle evolution of the growth rate spectrum on four flux surfaces $\Psi_N=0.95, 0.96, 0.97, 0.98$. 
Deepest inside the core, $\Psi_N =0.95$ sits just inside the shallow gradient core plasma when the pedestal is fully developed 
(corresponding to surface $b$ of Figure~\ref{fig:pedcartoon} below). MTMs dominate the spectrum in Figure~\ref{fig:specevol}(a), 
which broadens and increases in amplitude as locally $\beta$ rises through the MAST ELM cycle.
MTMs may limit the electron temperature gradient in this region of the plasma. We note that magnetic drifts are important to the microtearing mode \cite{APPLEGATE_PPCF2007}, and that $dT_e/dr$ may be more severely limited in equilibria with more unfavourable drifts.
Further out, Figure~\ref{fig:specevol}(c) shows a striking transition in $\gamma(k_y \rho_i)$, when the $\Psi_N=0.97$ surface joins the expanding pedestal at $t \sim 0.6$: the maximum growth rate, $\gamma^{\rm max}$, falls and then rises; and the dominant wavenumber bifurcates to a lower value. Inspection of the eigenfunctions reveals that MTMs, centred at $k_y \rho_i \sim 3$, dominate while $\Psi_N=0.97$ experiences core conditions, but that the MTMs are suppressed and supplanted by KBMs at $k_y \rho_i \sim 0.2$ when $dP/dr$ steepens to the higher pedestal value. The physics driving this transition is investigated below. Further into the pedestal at $\Psi_N=0.98$,
Figure~\ref{fig:specevol}(d) shows that KBMs are driven unstable earlier, but their growth rates fall later in the ELM cycle. KBMs remain the dominant modes, but for $t> 0.5$ increasing bootstrap current reduces the magnetic shear, which pushes the modes towards marginal stability.

How might the microinstabilities physically influence the evolution of the MAST pedestal? To deepen our understanding of the basic mechanisms, several artificial stability scans have been performed based around the key surfaces illustrated schematically in figure~\ref{fig:pedcartoon}.
\begin{figure}[h]
\setlength{\unitlength}{1cm}
\vspace*{-0.3cm}
\begin{center}
\includegraphics[scale=0.18]{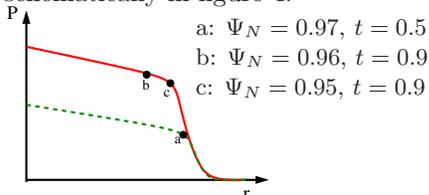}
\put(-1.,2.2){\small a: $\Psi_N=0.97$, $t=0.5$}
\put(-1.,1.8){\small b: $\Psi_N=0.96$, $t=0.9$}
\put(-1.,1.4){\small c: $\Psi_N=0.95$, $t=0.9$}
\hspace*{2cm}
\end{center}
\vspace*{-0.8cm}
\caption{\label{fig:pedcartoon} Schematic $P$ profiles from during the growth of the pedestal (dashed green) and when it is fully developed (solid red), indicating three reference surfaces for stability scans.}
\vspace*{-2mm}
\end{figure}

\begin{figure}
\vspace*{-0.9cm}
\setlength{\unitlength}{1cm}
\begin{center}
\hspace*{-1.2cm}
\includegraphics[scale=0.38]{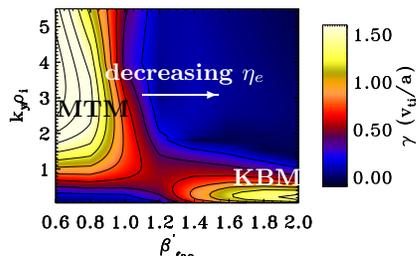}
\put(-5.15,2.0){\color{black} \bf \small MTM}
\put(-2.8,1.1){\color{white} \bf \small KBM}
\put(-4.5,2.5){\color{white} \bf \small decreasing $\eta_e$}
\put(-4.,2.3){\color{white} \vector(1,0){1.}}
\vspace*{-0.9cm}
\end{center}
\caption{\label{fig:pedadv} $\gamma(k_y \rho_i)$ scan around surface $a$ ($\Psi_N=0.97$ $t=0.5$) scaling $dP/dr$ and $R/L_n$ self-consistently. }
\vspace*{-2mm}
\end{figure}

\noindent
{\em Why does the pedestal expand inwards?--}
Artificial scans, around the reference equilibrium flux-surface $a$ in Figure~\ref{fig:pedcartoon}, 
have probed in more detail the mechanism by which the pedestal expands inwards. Surface $a$ ($\Psi_N=0.97$,$t=0.5$) sits just inside the shallow gradient core plasma, and will imminently be subsumed by the advancing pedestal. 

Figure~\ref{fig:pedadv} shows the growth rate spectrum from an equilibrium scan around surface $a$, multiplying the pressure gradient $\beta^{\prime}$ by a factor $\beta^{\prime}_{\rm fac}$ and scaling the density gradient $R/L_{n}$ and $\eta_e$ 
self-consistently at fixed $n_e$, $T_e$, $T_i$, and temperature gradients $R/{L_{T_e}}$ and  $R/{L_{T_i}}$. 
This scan closely resembles the evolution observed during the MAST ELM cycle, where the steepening pressure gradient is dominated by an increasing density gradient. 
The dominant $\gamma$ and $k_y$ fall as $\beta^{\prime}$ increases above the experimental value. 
The tearing parity mode is predominantly stabilised by the disruption of the relative phases between current and density perturbations arising from the increasing $R/L_n$ drive terms, and 
further stabilisation arises from more favourable drifts at higher $\beta^{\prime}$ \cite{BOURDELLE_NF2005,ROACH_PPCF1995}. 
This stabilisation of dominant MTMs reduces all transport associated with these modes. The quasilinear electron heat flux from the modes is larger than the particle flux, as is typical for MTMs.
At higher $R/L_{n}$, KBMs become dominant as their growth rates increase; KBMs transport heat and particles in all channels. 
The maximum $\beta^{\prime}$ at $\Psi_N=0.97$, from late in the ELM cycle, corresponds to $\beta^{\prime}_{\rm fac}=1.78$ which lies in the KBM region of figure~{\ref{fig:pedadv}}.

This scan suggests that: electron heat flux, and modest particle flux, initially fall as $dn_e/dr$ increases, due to MTMs becoming suppressed; and at higher $dn_e/dr$ all transport fluxes increase strongly when KBMs become dominant. This reduction in transport will allow the electron pressure and density pedestals to build up during the ELM cycle. The particle source from edge neutrals is likely to play a significant role in this evolution of $n_e$. 
This picture is qualitatively consistent with the inwards advance of the pedestal in MAST, whereby the shallow pressure gradient at $a$ in Figure~\ref{fig:pedcartoon} undergoes a rapid transition to the steeper value in the pedestal. The relative impact on the evolution of $n_e$ and $T_e$ pedestal profiles will be sensitive to edge sources of heat and particles, and most notably the particle source from cold neutrals. A transient increase in the particle source, e.g. from a pellet, may accelerate the development of the density and pressure pedestal. 

\noindent
{\em What limits the inward expansion?:--}
Surface $b$ of Figure~\ref{fig:pedcartoon} ($\Psi_N =0.95$, $t=0.9$) sits well within the shallow gradient core plasma when the pedestal is almost fully developed. $\beta$ would clearly have to increase on this surface if the pedestal were to advance further into the core by raising $dP_e/dr$ on surfaces where $\Psi_N >0.95$. This motivated an artificial scan centered on surface $b$, where $\beta$ and $n_e$ were scaled by $\beta_{\rm fac}$ consistently with the observed pedestal evolution in MAST. $\beta^{\prime}$ (which influences the magnetic drifts) was adjusted self-consistently for fixed $dn_e/dr$, $T$, $R/L_T$ and collision frequencies. The width of the unstable MTM spectrum was found to increase marginally with $\beta_{\rm fac}$.
\begin{figure}[h]
\vspace*{-1.1cm}
\setlength{\unitlength}{1cm}
\begin{center}
\includegraphics[scale=0.36]{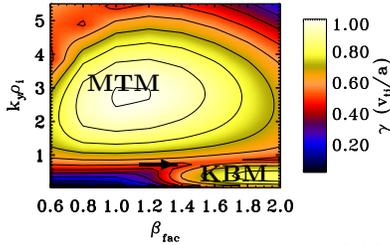}
\put(-4.4,2.1){\color{black} \bf \small MTM}
\put(-2.9,0.9){\color{black} \small \bf KBM}
\thicklines
\put(-3.7,1.12){\color{black} \vector(1,0){0.5}}
\vspace*{-0.9cm}
\end{center}
\caption{\label{fig:pedarrest} $\gamma$ spectrum scan around $\Psi_N=0.96$, $t=0.9$ (surface $c$) scaling $\beta$ and  $n_e$ as described in the text.}
\vspace*{-4mm}
\end{figure}

More interestingly, this scan was repeated on surface $c$ of Figure~\ref{fig:pedcartoon} ($\Psi_N=0.96$, $t=0.9$), 
which is further out than $b$, in the transition region between the shallow core and the steep pedestal pressure gradient just prior to the ELM. Figure~\ref{fig:pedarrest} indicates that this 
surface sits close to a critical value of $\beta$ where {\bf strongly growing KBMs and MTMs coexist over a broad spectrum}
and are equally dominant at the arrow, where $n \sim 25$. 
The crossing of this threshold may trigger a significant transport event in the edge plasma. 
$n=25$ coincides with the toroidal mode number of the 
peeling-ballooning mode found marginally unstable at the end of this ELM cycle {\cite{DICKINSON_PPCF2011}}, 
and is in the observed range for ELM filaments in MAST {\cite{KIRK_PPCF2011}}. \\
\noindent
{\em Conclusions:--} 
Plasma equilibria have been reconstructed from the spherical tokamak MAST, with sufficient resolution in time and space to capture plasma 
evolution during the short period between ELMs. Immediately after the ELM, steep pedestal 
pressure and density gradients form close to the separatrix, and these advance into the core.
Fully electromagnetic local linear gyrokinetic analysis reveals the dominant 
microinstabilities as a function of radius and time through the ELM cycle.  
This suggests a new physical picture of the
formation of this pedestal: steep pedestal pressure and
density gradients are limited by
KBMs; MTMs dominate and limit $R/L_{T_e}$ in the shallower
pressure gradient region of the core just inside the pedestal;
the pressure pedestal can propagate into the core because
increasing $R/L_n$ and $dP/dr$ stabilizes the MTMs until they
become supplanted by KBMs at a higher pressure gradient;
and meanwhile in the core MTMs become increasingly virulent as 
$\beta$ increases in the shallower pressure gradient
region. Furthermore, when the pedestal is nearest to being fully 
developed, the gradient transition region is close
to an instability threshold where MTMs and KBMs become simultaneously 
unstable with large growth rates over
a broad spectral range. Breaching this limit may trigger
signiﬁcant change in the edge transport.
This picture should be compared with data from a broader range of H-mode plasmas and from turbulence diagnostics, 
and extended to predict transport timescales using nonlinear gyrokinetic fluxes and realistic sources.\\
\noindent
{\em Acknowledgements:--}
This work was funded by the RCUK Energy Programme under grant EP/I501045 and by the European Communities under the contract of Association between
EURATOM and CCFE. The views and opinions expressed do not necessarily reflect those of the European Commission. Authors acknowledge stimulating 
discussions with J W Connor, S C Cowley and R J Akers, and access to the HECToR supercomputer through EPSRC grant EP/H002081/1.
\bibliographystyle{aip}

\end{document}